\newcommand{\hmpc}{\mbox{ } {h}^{-1} \rm{ Mpc}}
\newcommand{\hkpc}{\mbox{ } {h}^{-1} \rm{ kpc}}
\newcommand{\hmsol}{\mbox{ } {h}^{-1}  {M}_{\odot}}
\newcommand{\msol}{\mbox{ } {M}_{\odot}}
\newcommand{\lcdm}{$\Lambda$CDM }
\newcommand{\msig}{{M}_{\rm bh}-\sigma}
\newcommand{\mhalo}{{M}_{\rm bh}-{M}_{\rm tot}}
\newcommand{\mbh}{{M}_{\rm bh}}

\documentclass[floatfix,iop]{emulateapj}
\usepackage{epsfig,subfigure}

\bibpunct[; ]{(}{)}{;}{a}{}{,}

\begin{document}
\title{Examining subgrid models of supermassive black holes in cosmological simulation}

\author{P.~M.~Sutter} \email{psutter2@illinois.edu}
\affil{Department of Physics,
	     University of Illinois at Urbana-Champaign,
            Urbana, IL 61801-3080}

\and

\author{P.~M.~Ricker} \email{pmricker@illinois.edu}
\affil{Department of Astronomy,
       University of Illinois at Urbana-Champaign,
             Urbana, IL 61801\\
		National Center for Supercomputing Applications,
      University of Illinois at Urbana-Champaign,
            Urbana, IL 61801}
            
\begin{abstract}
While supermassive black holes (SMBHs) play an important role
in galaxy and cluster evolution, at present they can only be included
in large-scale cosmological simulation via subgrid techniques.
However, these subgrid models have not been studied in a
systematic fashion.
Using a newly developed fast, parallel spherical
overdensity halo finder built into
the simulation code FLASH, we perform a suite of
dark matter-only cosmological simulations to study the effects of
subgrid model choice on relations between SMBH
mass and dark matter halo mass and velocity dispersion.
We examine three aspects of SMBH subgrid models: the choice of initial
black hole seed mass, the test for merging two black holes, and
the frequency of applying the subgrid model. We also examine the
role that merging can play in determining the relations, ignoring 
the complicating effects of SMBH-driven accretion and feedback. 
We find that the choice 
of subgrid model can dramatically affect the black hole merger rate, 
the cosmic SMBH mass density, and the low-redshift relations to halo 
properties. We also find that it is possible to reproduce observations 
of the low-redshift relations without accretion and feedback, 
depending on the choice of subgrid model.  
\end{abstract}
\maketitle

\section{Introduction}
\label{sec:introduction}
Supermassive black holes (SMBHs) play a number of 
roles in the evolution and dynamics of galaxies and clusters of galaxies. 
These black holes, with masses of at least a million solar masses, 
and their associated accretion disks
drive quasars at high redshift~\citep{fan_evolution_2006}, 
regulate star formation in galaxies~\citep{hopkins_mergers_2010}, 
inject thermal energy into the intracluster medium 
via powerful jets~\citep{gu_bulk_2009}, 
and may even play a large role in establishing kiloparsec-scale 
microgauss magnetic fields in clusters~\citep{carilli_c_2002}.

Of particular interest are the correlations discovered between 
SMBH mass and other observed quantities of galaxies.
The first discovered relationship, between SMBH mass 
and bulge stellar luminosity~\citep{magorrian_demography_1998},
was intriguing but suffered from large scatter. Subsequent searches 
found a tight correlation between SMBH mass and bulge 
velocity dispersion 
$\sigma$~\citep{tremaine_slope_2002, gultekin_m-sigma_2009}, 
although comparable improvements have also been made in the 
correlations to stellar luminosity~\citep{graham_black_2007}. 
Most recently, black hole 
mass has been linked to dark matter halo mass ${M}_{\rm tot}$, both 
indirectly by measuring galactic circular 
velocities (e.g.,~\citep{ferrarese_beyondbulge:fundamental_2002};
~\citep{baes_observational_2003})
and by direct estimates of halo mass via gravitational 
lensing~\citep{bandara_relationship_2009}. 
The latter measurements agree well with 
theoretical predictions of the relationship between 
halo virial mass and galactic circular velocity~\citep{croton_simple_2009}.

These relationships imply a correlation between the growth 
of SMBHs and that of their host galaxies. Since structures form in the 
universe via hierarchical clustering 
of smaller objects~\citep{baugh_primerhierarchical_2006},  
black holes carried along with their hosts should tend to merge 
as well~\citep{hopkins_black_2005}. 
This may provide a simple and direct scaling between 
halo and black hole mass, especially at 
low redshift~\citep{volonteri_assembly_2003}. However, 
feedback processes driven by accretion disk systems may also 
contribute to the observed correlations: greater accretion 
can lead to larger feedback events, thereby reducing the accretion 
rate and coupling the black hole mass to the 
surrounding system~\citep{cattaneo_agn_2007}.

Attempts to explain the observed relations via cosmological simulations
require large volumes (to gather enough objects) and high 
resolution (to capture galaxy-sized structures). However, 
even with current computing resources, large-volume simulations 
are unable to capture all the intricate physical processes that 
dominate in the formation, merging, and feedback of black holes. 
Hence, these processes must be added in post-processing 
as semi-analytic models (e.g., ~\citealt{micic_supermassive_2007}) 
or included in-situ as subgrid models 
(e.g.,~\citealt{booth_cosmological_2009}
;~\citealt{sijacki_unified_2007}
;~\citealt{di_matteo_direct_2008}). 

Many processes have been proposed to explain the formation 
of seed SMBHs in the early universe, 
including remnants of Population III 
stars~\citep{madau_massive_2001, wise_number_2005}, 
direct collapse of gas in 
central bulges~\citep{koushiappas_massive_2004, begelman_formation_2006}, 
and merging of smaller black holes~\citep{islam_massive_2004}. 
Merging black holes are difficult systems to study, since 
they interact with their gaseous environment~\citep{mayer_rapid_2007}, 
emit gravitational radiation~\citep{sesana_lowfrequency_2004},
and can suffer kicks due to merging~\citep{baker_modeling_2008}.
Additionally, the self-regulating feedback processes 
emerging from accretion onto black holes and the 
subsequent formation of jets and bubbles are not 
fully understood~\citep{vernaleo_agn_2006}, 
especially when considering the 
effects of magnetic fields~\citep{ruszkowski_impact_2007}
and 
turbulence~\citep{brggen_self-regulation_2009}. Consequently, subgrid models 
must make many simplifying assumptions in treating these processes.

The variety of plausible scenarios for forming and merging 
black holes and applying feedback processes allows modelers great 
latitude in developing and adjusting models to fit observations. 
Universally, all aspects of the formation and evolution of SMBHs 
are combined in the same simulation. However, we believe that 
subgrid models of the initial seeding and merging of SMBHs (which are 
linked in subgrid models to 
the properties of the surrounding dark matter) should be 
separated from models of accretion and feedback (which depend on the 
local gas physics). This way, we can better understand the role 
that merging alone plays in developing the $\msig$ and $\mhalo$ relations 
and the effects of changing subgrid models on those same relations.

Thus, in this paper we examine dark matter-only simulations of the growth 
of structure in a cosmological volume including subgrid models 
to track the formation and merging of SMBHs.
By comparing the results of several plausible scenarios for models
against observed relations, we will determine how much of those relations 
is due to low-redshift and large-scale evolution of SMBHs, and how these 
models may affect the final outcomes, independent of any gas accretion 
or feedback.

To follow our dark matter halos, we have developed a new, parallel, fast
halo finder built directly into the simulation code FLASH v2.5.
FLASH is an adaptive-mesh refinement (AMR)
code for astrophysics and cosmology~\citep{fryxell_flash:adaptive_2000}.
FLASH solves the N-body potential
problem with a particle-mesh multigrid fast Fourier transform
 method~\citep{ricker_direct_2008}.
It uses smoothed cloud-in-cell mapping 
(Ricker et al. in preparation 2010)
for interpolating between
the mesh and particles~\citep{hockney_computer_1988} and
a second-order leapfrog integration scheme
for variable time step particle advancement.

In the following section we discuss the precision and valid mass 
ranges for our new halo finder. In Section~\ref{sec:subgrid} we outline 
the numerical aspects of our approach and the black hole 
formation and merging subgrid models 
employed. Finally in Section~\ref{sec:analysis} we compare our 
results to observations of the $\msig$ and $\mhalo$ 
relations to test the validity of the models. Additionally, 
we provide a discussion and analysis of the performance 
and parallel scalability of our halo finder in the Appendix.

\section{The Halo Finding Method}
\label{sec:haloFinding}
We base the halo finder in FLASH on a spherical-overdensity (SO) 
technique. Throughout, we will identify our new halo finder by 
``pSO'', for parallel spherical overdensity. 
In this approach, halos are defined by spherical regions 
within which the mean density is greater than some defined threshold. 
We begin by mapping particles onto the simulation 
mesh with smoothed cloud-in-cell mapping 
and identifying peaks by finding zones with 
densities greater then all surrounding zones 
and greater than 
$\Delta_{\rm peak} \rho_{\rm crit}$. Here and throughout, $\rho_{\rm crit}$ 
refers to the comoving critical density of the universe,
\begin{equation}
  \rho_{\rm crit} = \frac{3 H_0^2}{8 \pi G} \left[ 
                    \Omega_{M,0} 
                    + \Omega_{\Lambda,0} (1+z)^{-3} \right],
\label{eq:critDen}
\end{equation}
where subscripts of $0$ here and throughout refer to present-day values.
The zone midpoints serve as centers of potential halos. 
Using a binary search, we compare the average density within 
the current search radius to $\Delta_{\rm search} \rho_{\rm crit}$, selecting 
new search radii appropriately. Initially, search radii are doubled 
until the enclosed density is below the threshold. Only then 
does the binary procedure begin. 
When two successive 
search radii differ by no more than a chosen amount, defined 
by the parameter $\Delta R_{\rm stop}$, we stop the search. 
If, during the search, a radius falls below a cutoff value, 
$\Delta R_{\rm small}$, we the abort 
the search and disregard the halo. Finally, we remove 
any finished halos that have 
radii smaller than $\Delta R_{\rm min}$ from the catalog.
We also remove satellite halos whose centers fall within 
the radii of larger companions. This leads to a more consistent mass 
function and conserves halo mass~\citep{white_mass_2002}.
We may also optionally remove satellite halos which intersect larger 
neighbors. While this must be done to strictly conserve halo mass, 
many authors include these satellite halos, since it is sometimes 
useful to identify satellite structures (see the comparison 
in~\citet{evrard_virial_2008} for a discussion of this decision).
Table~\ref{tab:hfParams} lists the parameters controlling our halo 
finder and our chosen values. 
In the table and throughout, $\Delta x$ is the grid resolution. 
Note that for simulations in which the halos may span an adaptively refined 
region, $\Delta x$ will refer to the highest-resolution uniformly 
refined mesh level.

\begin{table}
  \centering
  \caption{Parameters controlling the pSO halo finder in FLASH.}
  \begin{tabular}{ccc}
    \hline
    \hline
  Parameter & Description  & Value \\
  \hline
  $\Delta_{\rm peak}$   & Overdensity for identifying a halo center   & $200 \rho_{\rm crit}$  \\
  $\Delta_{\rm search}$ & Overdensity for defining a halo & $200 \rho_{\rm crit}$ \\
  $\Delta R_{\rm stop}$   & Criterion for completing a radius search   & $0.2 \Delta x$  \\
  $\Delta R_{\rm small}$   & Criterion for aborting a radius search  & $0.5 \Delta x$  \\
  $\Delta R_{\rm min}$   & Minimum resolvable halo radius  & $1.0 \Delta x$  \\
  \tableline
  \end{tabular}
  \tablecomments{$\Delta x$ is the uniform grid resolution.}
  \label{tab:hfParams}
\end{table}

For this work, we chose 
$\Delta_{\rm peak} = \Delta_{\rm search} = 200$.
However, other values may be chosen
for other uses of the halo finder. For example, $\Delta_{\rm search} = 500$ 
would be appropriate for generating mock observations of X-ray cores 
for comparison to observations~\citep{evrard_mass_1996}, 
while a smaller value might be 
useful for triggering refinement.

To evaluate our halo finder, we performed dark matter-only simulations in a 
cubic $128 \hmpc$ 
box with $256^3$, $512^3$, and $1024^3$ zones, giving 
resolutions of 
$\Delta x = 500$, $250$, and $125 \hkpc$ respectively. 
All runs used $256^3$ particles, 
giving a mass resolution of $1.3 \times 10^{10} \msol$. 
For these tests, we chose cosmological parameter values of 
$\Omega_{M,0} = 0.26$, $\Omega_{\Lambda,0}=0.74$, and
 $H_0 = 100 h = 71 \mbox{ km s}^{-1} \mbox{Mpc}^{-1}$ for comparison to
 runs used in previous works.
We compared our pSO halos 
against spherically overdense regions drawn from a friends-of-friends (FOF)
halo finder with a linking length of $0.2$ 
(see \citealt{lukic_structure_2009} 
for an analysis of SO halos drawn from an FOF 
catalog). 
The halos drawn from the FOF catalog used the 
most-linked particle as the halo center. Halo radii in the FOF catalog 
were determined by starting 
with a large radius and moving inwards at very small increments 
(much smaller than the $\Delta R_{\rm stop}$ used in pSO)
until the interior density 
exceeded the same threshold as used by the pSO halo finder. This approach 
is more precise, but much slower, than the binary search technique used 
in pSO. We use a linear search here since with such high precision 
a binary technique may not complete.

We thus have several possible sources of differences between the SO
and FOF halo catalogs: (1) pSO chooses the zone midpoints as the 
halo centers, while FOF halos use the most-linked particles, (2) the 
incremental search radii when finding spherical overdense
 regions in the FOF catalog 
are much smaller than those used in pSO, (3) FOF tends to 
find more small halos than the grid-based peak finding in pSO, since smaller 
halos may be irregularly shaped, and (4) FOF will tend to bridge two 
nearby halos, even if they have distinct spherically overdense regions.

\begin{figure}
  \centering
  \includegraphics[width=\columnwidth]{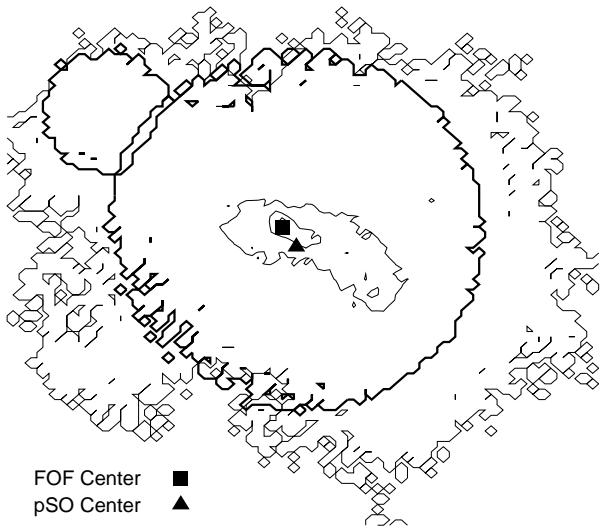}
  \caption{Demonstration of different choices of halo center 
           due to recent merging. 
           The thin lines indicate projected density contours 
           of the FOF halo. The interior contour lines define  
           thresholds of 50 and 200 particles per zone, while the
           outermost contour shows a threshold of 1 particle per zone
            (i.e., the boundary of the FOF halo). 
           The two thick circles show the boundaries
           of the spherically overdense regions identified by pSO.
           The 
           separation of the larger pSO 
           halo center (circle) and the FOF halo center (square)
           is roughly one half a zone.}
\label{fig:halos}
\end{figure}

Figure~\ref{fig:halos} demonstrates both the problem of selecting a halo 
center and that of pSO counting more satellite halos than FOF.
Shown is a $\sim 10^{14} \hmsol$ 
halo drawn from the run with $250 \hkpc$ resolution. 
This halo is undergoing a merger with a 
smaller satellite, and thus 
FOF is bridging two distinct spherical regions into a single halo. 
In agreement with the findings of~\citet{evrard_virial_2008}, 
we find roughly $\sim 15\%$ of our halos are satellites.
While the most-linked particle approximates the potential minimum well, 
the pSO halo center, which is simply the zone midpoint, is within a 
zone radius of this point.
The bridging effect leaves the smaller pSO halo unmatched.

We match halos in the pSO and FOF catalogs by finding intersections. Any 
halos from the two catalogs that overlap such that they contain each other's 
centers are considered a match. Due to the ambiguity of halos with very few 
particles, we only compare halos with more than $100$ particles. 
With this procedure alone, about 15\% 
of the pSO halos remained unmatched. To determine if this was due to 
the FOF bridging effect, we removed from consideration any pSO halo 
that was within a linking length of an already matched FOF halo. This 
accounted for all the unmatched pSO halos. 

\begin{figure}
  \centering
  \includegraphics[width=\columnwidth]{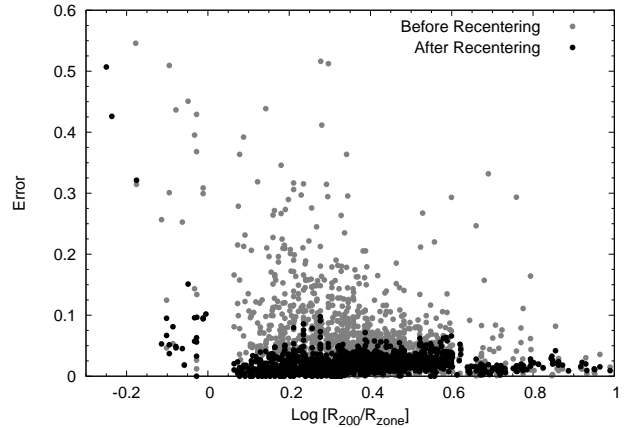}
  \caption{Error in matched halos before and after recentering FOF
           halos on pSO centers, as a function of halo radius. 
           The error is defined in the text.}
\label{fig:recenter}
\end{figure}

To account for the effects of choice of halo center, we re-computed 
overdensities in the FOF catalog using the potential minimum halo centers 
found in the matching pSO catalog. Note that we found that 
the differences between 
halo centers were always smaller than a zone spacing. 
There may be some cases, however, where a large asymmetry will 
separate the most-linked particle and the maximum density peak 
further than a zone spacing, especially at very high resolutions. 
Here, asymmetries will manifest themselves in a higher relative error 
between matched halos.
Figure~\ref{fig:recenter} shows 
errors in the matched halos in the $125 \hkpc$ catalog at $z=0$
before and after recentering. We define the errors as 
the difference in mass divided by the sum of the $1\sigma$ 
uncertainties in the halo mass:
\begin{equation}
  E \equiv \frac{\left| M_1 - M_2 \right|}
                {\sigma_{{\rm M}_1} + \sigma_{{\rm M}_2} }.
\label{eq:error}
\end{equation}
We estimate the halo mass uncertainty 
by assuming an uncertainty in the halo radius of $0.5 \Delta x$. Thus, 
assuming constant halo density, the mass uncertainty is
\begin{equation}
  \sigma_{\rm M} = \frac{3}{2} \frac{M}{R} \Delta x \propto M^{2/3},
\label{eq:deltaM}
\end{equation}
where $M$ and $R$ refer to the halo mass and radius, respectively. 
For halos above $10^{14} \msol$ in mass, this leads to 
$\sigma_{\rm M} / M \approx 0.1$, 
which agrees with the estimates of~\citet{bhattacharya_mass_2010}.
While most matched halos have 
small error, a few differ by as much as $0.7$, especially 
at lower masses. However, most of this 
is due to the bridging effect's having moved the most-linked particle away from 
the potential minimum. After recentering, errors for all halos larger 
than a zone radius drop to below $0.1$.  
The small gap in halo sizes near $R_{\rm 200}/R_{\rm zone} = 0$ 
is due to our choice of 
$\Delta R_{\rm stop}$ and the binary search procedure.

Also from Figure~\ref{fig:recenter} we find that even after recentering,
halos smaller than a zone do not match well to FOF halos. Thus 
we set the parameter $\Delta R_{\rm min}$ to $1.0 \Delta x$ and reject 
any halo smaller than this.
Given a resolution, this sets our minimum 
resolvable halo mass. 
Note that this criterion is consistent with the estimate for minimum 
resolvable mass for FOF halos as defined
by~\citet{lukic_halo_2007}:
\begin{equation}
n_{{\rm min}} = \frac{\Delta (1.61 n_p/n_g)^3}{\Omega_{M,0} (1+z)^3 } 
\left[ \Omega_{M,0}(1+z)^3 + \Omega_{\Lambda,0} \right],
\label{eq:minMass}
\end{equation}
where $n_g$ and $n_p$ are the number of zones and particles per side, 
respectively, and $\Delta=200$ is our chosen overdensity.

\begin{figure}
  \centering
  \includegraphics[width=\columnwidth]{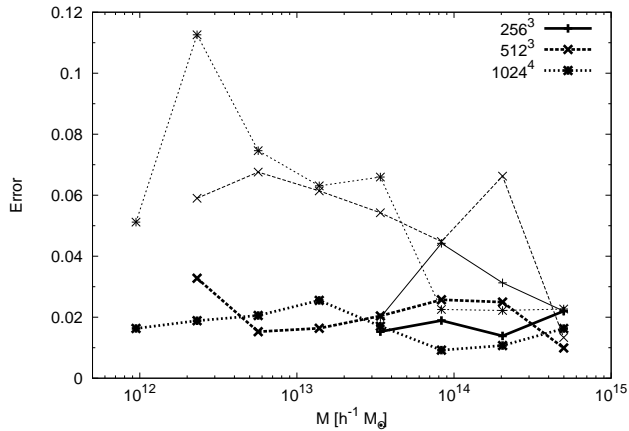}
  \caption{Error in matched halos at different simulation resolutions. 
           Bold lines denote average error, while thin lines describe 
           maximum error for each mass bin. The error is defined in the text, 
           and is computed after recentering.}
\label{fig:errorRes}
\end{figure}

Figure~\ref{fig:errorRes} shows maximum and average errors 
after recentering for each of $10$ mass bins at $z=0$ 
for each simulation resolution. With our approach, we 
are able to maintain average errors of less than $0.04$ for all 
resolvable halos, while a small number ($< 5\%$) of halos have 
maximum errors of up to $0.12$. Although the average error stays consistent 
across all masses, the maximum error varies as much as $0.04$ for adjacent 
bins, and it rises with decreasing mass. This behavior is due to a 
small number of irregularly-shaped halos. Also, since smaller halos 
have smaller uncertainties, the errors tend to become larger with 
smaller halo mass.

\begin{figure}
  \centering
  \includegraphics[width=\columnwidth]{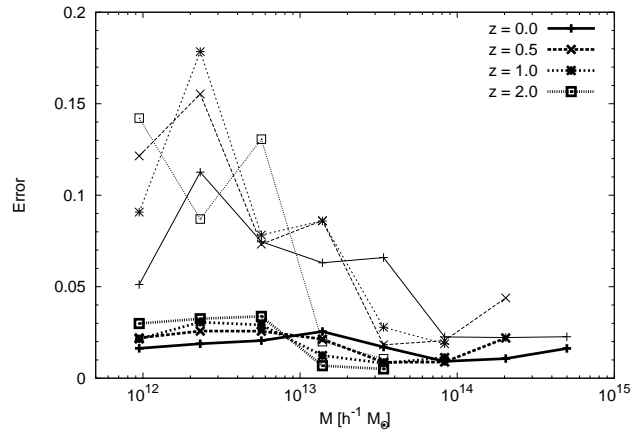}
  \caption{Error in matched halos at different redshifts for 
           run with $125 \hkpc$  resolution. Bold lines show average error 
           for each bin, while thin lines denote maximum error. The error
           is defined in the text, and is computed after recentering.}
\label{fig:errorRed}
\end{figure}

Figure~\ref{fig:errorRed} 
shows errors after recentering 
at various redshifts in the $\Delta x = 125 \hkpc$ run. 
Across all resolvable masses and redshifts from $z=0$ to $2$, 
we are able to maintain average errors less than $0.05$. 
However, maximum errors in the smallest mass bins reach as 
high as $0.18$. Again, the variability of the maximum errors 
is due to a small number of halos.

By adjusting $\Delta R_{\rm stop}$, we are able to reduce both the average 
 and maximum errors. 
However, at smaller values we found that the binary search procedure 
had difficulty converging on a value for some halos, 
and the halo finder ran for an unacceptable amount of time. Larger 
values produced unacceptably high errors.
For the value we chose, 
after accounting for different choices of halo center, the halos 
produced by our new pSO halo finder are statistically indistinguishable 
from matched halos drawn from a traditional FOF halo finder. 

\begin{figure}
  \centering
  \includegraphics[width=\columnwidth]{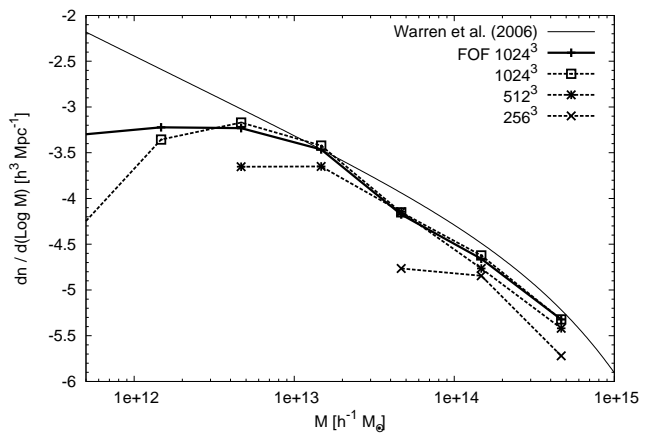}
  \caption{Mass functions at different resolutions compared against 
           halos drawn from the $125 \hkpc$ resolution FOF catalog and 
           against the mass function from~\citet{warren_precision_2006}.}
\label{fig:massFunc}
\end{figure}

Figure~\ref{fig:massFunc} shows the mass function for the pSO halo 
catalogs at the three resolutions compared to SO halos 
drawn from an FOF catalog with a resolution of $125 \hkpc$ and compared 
to the mass function obtained by~\citet{warren_precision_2006}. 
As expected, pSO captures more halos as the 
resolution increases. 
However, pSO captures fewer halos near 
the resolvability limit than FOF does. 
The procedure for mapping and smoothing tends to lower the central 
density, especially with halos near the resolvability limit.
Thus our criterion for selecting a peak, $\Delta R_{\rm peak} = 200$, may 
be too stringent. We found that lowering this value captures more halos, 
but at the expense of identifying too many potential halos that end 
up below the resolvability limit. Even at values as low as $100$, we still 
could not replicate the FOF mass function at these masses, 
while adding roughly four times as many 
candidate halos that were ultimately rejected. 

Finally, to speed up processing we may increase $\Delta R_{\rm small}$. 
However, 
the binary search may briefly fall below this threshold before settling on a 
larger, and correct, radius. Thus we want to pick the largest possible 
value that does not cause us to remove resolved halos. We found that at any 
values above $0.5 \Delta x$ we began to reject valid halos before 
their search had completed.

\section{Subgrid Models}
\label{sec:subgrid}
We identified three areas in which subgrid models may differ: 
the initial mass of seed black holes, the prescription for 
determining if two black holes merge, and the frequency 
of finding halos, creating black holes, and checking for mergers.
For each of these aspects we study two possibilities: a 
``pessimistic'' and an ``optimistic'' scenario. This leads
to a total of eight combinations of models. 

For the initial mass of seed black holes, authors who have performed 
calculations similar to ours 
(e.g.,~\citealt{booth_cosmological_2009}
;~\citealt{sijacki_unified_2007}
;~\citealt{di_matteo_direct_2008}) 
 typically choose a constant seed mass of $10^{5} \msol$, and this is 
the value we will choose for this model.
The value chosen is typically motivated by a desire not to violate constraints 
on the observed black hole mass 
density~\citep{shankar_supermassive_2004}, especially when black holes
are allowed to accrete a significant portion of their final mass. 

However, this approach may have several weaknesses. First, black holes 
with masses as high as $10^9 \msol$ are inferred to exist 
from quasar activity at $z \sim 6$~\citep{fan_evolution_2006}. 
Second, depending on the frequency of 
halo finding, when halos are initially detected they may not have 
identical masses, and if whatever physical processes which 
create the observed relations are already present, then 
the seed black holes should scale with halo mass. Finally, at lower 
redshifts the halo finding algorithm may spuriously merge two halos.
If we instantly merge black holes and in the following step the 
halo finder identifies two separate halos, one will be without 
its black hole. It may be inappropriate to re-seed the halo 
with a low-mass black hole. 

Alternatively, we may derive the seed black hole mass from a form 
of the observed $\msig$ relation.
This approach may seem tautological; however, accretion and feedback 
processes may dominate at high redshift and at scales below which 
we can resolve~\citep{shankar_demography_2009, merloni_cosmic_2010}. 
The observed $\msig$ relation at $z=0$ 
may then simply be a consequence of pure merging at late times 
and large scales. Also, this approach may be useful in testing 
other SMBH-related processes and relations in groups and clusters.  
However, this approach is complicated by the fact that 
calculations of $\sigma$ directly in simulations suffer from 
high scatter, especially at high redshift when halos first 
form and at low redshift when satellite halos may be contaminated 
by hot particles belonging to a larger neighbor. 
Fortunately, we may seed halos by relating the black hole 
mass directly to the halo mass through the observed 
relation~\citep{bandara_relationship_2009}:
\begin{equation}
\log \left( \frac{{M}_{\rm bh}}{\msol} \right)  
         = (8.18 \pm 0.11) 
         + (1.55 \pm 0.31)
         \log\left( \frac{{M}_{\rm tot}}{10^{13} \msol} \right).
\label{eq:mhalo}
\end{equation}
This relation comes from observations at $z=0.2$ of bulge 
velocity dispersions and halo masses 
using gravitational lensing. 
These measurements are linked to black hole mass by assuming the 
$\msig$ relation of~\citet{gultekin_m-sigma_2009}:
\begin{equation}
\log \left( \frac{{M}_{\rm bh}}{\msol} \right)  
         = (8.12 \pm 0.11) 
         + (4.24 \pm 0.31)
         \log \left( \frac{\sigma}{\sigma_0} \right),
\label{eq:msigma}
\end{equation}
where $\sigma_0 = 200 \mbox{ km s}^{-1}$. 
We will assume the relation in Eq.(\ref{eq:mhalo}) holds to high 
redshift.  
 
\begin{figure}
  \centering
  \includegraphics[width=\columnwidth]{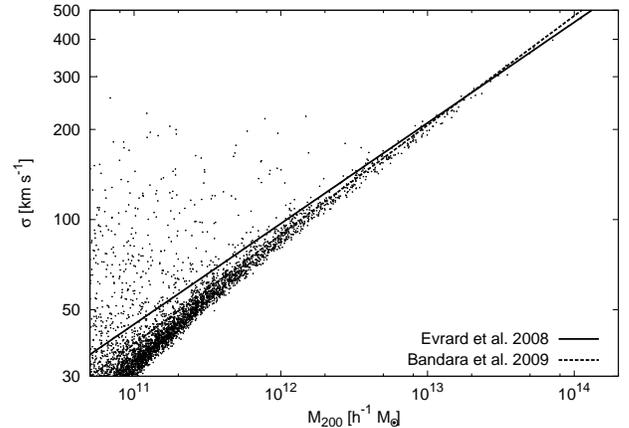}
  \caption{$\sigma_{\rm DM}$ as a function of halo mass at redshift $z=0.2$, 
           compared to observations~\citep{bandara_relationship_2009} 
           and best fits from simulations~\citep{evrard_virial_2008}.
           The latter is an extrapolation below $\sim 10^{14} \msol$.}
\label{fig:msigma200}
\end{figure}

Figure~\ref{fig:msigma200} shows velocity dispersion as measured in one of 
our runs (detailed below) as a function of mass at $z=0.2$. While 
we agree with observations and the best fit from a suite of 
simulations described by~\cite{evrard_virial_2008}, there is significant 
scatter, especially at low mass. Note that since we are performing 
 dark matter-only simulations, we do not directly measure the velocity 
dispersion of the galactic central bulge; instead we measure that 
of the dark matter in the entire halo, defined by:
\begin{equation}
 \sigma_{\rm DM}^2 = a(t)^2 \frac{1}{3 N_p} \sum_{i=1}^{N_p} \sum_{j=1}^{3} 
                  \left| v_{i,j} - \bar v_{j} \right|^2,
\label{eq:sigma}
\end{equation}
where $a(t)$ is the scale factor, 
$N_p$ is the number of particles in the halo, $v_{i,j}$ is the 
$j{\rm th}$ velocity component of the $i{\rm th}$ particle, and 
$\bar v_j$ is the $j{\rm th}$ component of the center-of-mass velocity.
Note that velocities in the simulation are comoving quantities. 
This plot and other results (as summarized by~\citealt{croton_simple_2009}) 
indicate that $\sigma_{\rm DM}$ and $\sigma_{\rm bulge}$ agree to within a 
factor of order unity, so we may substitute one for the other.

Next, we may change the prescription for testing 
black hole mergers.
Most approaches to studying the growth of SMBHs rely
on halo merger histories derived from 
simulations~\citep{micic_supermassive_2007} or 
Press-Schechter 
models~\citep{menou_merger_2001, volonteri_evolution_2008}. 
These approaches assume that 
each halo contains a single SMBH, and whenever two halos merge 
their respective black holes instantly merge as well. 
While this approach is highly optimistic, it does provide an 
upper limit to predicted merger rates, an important constraint 
on upcoming gravitational wave 
experiments~\citep{amaro-seoane_topical_2007}.
It is also easy to implement and provides a simple way 
to compare results from simulations to approaches based 
on merger trees. However, this approach can lead to spurious 
black hole mergers, since halos may temporarily intersect without 
merging. We will include this scenario in our study.

Subgrid models used directly in 
simulations~\citep[e.g.,][]{booth_cosmological_2009} have used a more 
sophisticated approach in merging black holes.  
In these simulations, SMBHs only merge 
when they are within some defined distance (for Lagrangian codes, 
this is typically the softening length) and have relative 
velocities below some threshold. This is to avoid merging black holes 
that are only passing by each other and are not part of a true 
coalescing system. 
The velocity threshold varies by author, but it is usually taken to be 
the local gas sound speed or the circular velocity near the 
larger black hole of the merging pair. 
However, we found this test to be overly restrictive; 
without gas to slow down black holes in cluster cores,
merging times 
inferred from these criteria are larger than the Hubble time and hence 
almost no black holes merge using this model. Since 
observations indicate that black holes 
do merge~\citep{merritt_massive_2005}, we must apply a 
less stringent test.

Simulations of merging galaxies suggest that the time for black 
holes to move from kiloparsec to parsec scales is typically 
$\sim 10$~${\rm Myr}$~\citep{dotti_supermassive_2007}. 
This assumption was used in the merger-tree analysis 
of~\citet{micic_black_2008}. When applying this test we will ignore the effects 
of gravitational recoil~\citep{bogdanovic_alignment_2007} 
and the ``final parsec problem''~\citep{berczik_efficient_2006}.
Thus, in this merging test we will only merge black holes 
if they are both within the same halo and 
are within two grid zones of each other,
 and their relative velocity times $10$~${\rm Myr}$ is less than that same 
distance. This approach allows black holes to pass near each other without 
merging if they are not part of a truly merging system and accounts 
for the time needed for the black holes to merge below our resolvable 
scales.  

Finally, we may alter the frequency of performing our subgrid analysis. 
One common approach is to check for new black holes and allow mergers such that 
the interval between successive checks is evenly spaced in the log 
of the expansion factor.~\citet{booth_cosmological_2009} employ the 
shortest such interval, such that $a_{\rm next} = 1.02a_{\rm current}$.
This approach requires roughly $1/6$ the number of halo searches 
compared to
searching every time step.
Since our halo finder is designed to be inexpensive, we may
perform checks at every time step. This may cause spurious 
formation and merging of black holes, especially with small 
halos near the resolvability limit. However, with infrequent checks 
we may miss the formation and merging of new small halos, underestimating 
both the amount of black hole mass and the merger rate in the simulation.
We will study checking both every time step and at an interval 
of $\Delta \log a = \log 1.02$.

Table~\ref{tab:models} summarizes the aspects of the subgrid model 
we are studying, our choices for modifying each aspect, and a 
shortened name that we will use to identify the models in 
plots and tables. 
For example, 
a model that uses a constant initial seed mass, merges 
black holes instantly, and performs a check at every time step 
would be identified by ``con,halo,dt''.
We note that the most commonly used model in the literature 
uses a combination of uniform initial mass, 
velocity tests at every time step for merging, and seeding 
new black holes evenly in log expansion factor. For our study, 
we have combined merging tests and seeding in the same step. 
This is closer to the approach used in merger tree analysis, where 
analysis can only take place on the available halo catalogs.
\begin{table}
  \centering
  \caption{Aspects of Subgrid Models Used in the Creation and 
           Merging of SMBHs.}
  \begin{tabular}{ccc}
    \hline
    \hline
  Aspect & Model  & Short Name \\
  \hline
  Seed mass & Constant  & con \\
            & $\msig$ relation  & m-s \\
  Merging strategy & Instantly on halo merger  & halo \\
            & Distance and velocity test  & prox \\
  Frequency & Every time step  & dt \\
            & Evenly spaced in $\log(a)$  & log \\
  \hline
  \end{tabular}
  \label{tab:models}
\end{table}

\section{Comparison of Models}
\label{sec:analysis}
For all calculations,
we used concordance parameter values of
 $\Omega_{M,0} = 0.238$, $\Omega_{\Lambda,0}=0.762$, and
 $H_0 = 100 h = 73.0 \mbox{ km s}^{-1} \mbox{ Mpc}^{-1}$.
 All runs took place in a three-dimensional box
measuring $50 \hmpc$ per
side with $512^3$ particles and $1024^3$ zones per side, 
giving a mass resolution of $6.15 \times 10^7 \hmsol$ and spatial 
resolution of $48.8 \hkpc$.
There was no refinement of grid spacing.
All simulations used the same initial conditions:
unperturbed particle positions were situated on a grid,
and the initial velocities and positions were perturbed
using Gaussian fluctuations normalized to
$\sigma_8 = 0.74$. We assumed $P(k)$ from a \lcdm cosmology.
We used the GRAFIC2 code~\citep{bertschinger_multiscale_2001}
to generate these initial conditions.
All computations started at a redshift of $z=56.8$.

Using our halo finder, our minimum resolvable halo 
mass is $\sim 10^{10} \msol$. To help avoid spurious mergers, 
we did not include satellite halos (halos that intersect a 
larger neighbor) in the halo catalog 
used by the seeding and merging models. 
We seed black holes in any resolvable halo by creating a 
black hole particle in the simulation with a dynamical mass 
equal to the mass of the black hole and velocity equal to the 
center-of-mass velocity of the parent halo. When we merge black holes, 
we remove the smaller of the pair and add its mass to the larger of 
the pair. The larger SMBH maintains its position and velocity. 
We do not re-center black holes on halo potential minima. 

While re-positioning black holes at potential minima
reduces ejections due to scattering, this process 
may artificially promote merger rates. 
With our SMBH setup, we do indeed see a large 
ejection rate. Fortunately this is not an issue for our analysis: 
the vast majority of ejections are of un-merged black holes from 
low-mass halos at high redshift. 
Consequently, small black holes are simply replaced 
by another small black hole, and the merger rates and relations remain 
unaffected. Obviously, this would pose a problem for models 
that include accretion. The ideal solution in this case would be to 
initialize on halos well above the minimum resolvable halo mass and 
to adequately smooth the gravitational potential near the SMBH so 
that it does not experience significant two-body effects.
Also, by using the 
black hole mass as the dynamical mass, we may underestimate the 
dynamical friction. However, we have found that almost all black holes 
lie within a zone of their host's potential minimum, so this does not 
affect the merger rate.

\begin{figure*}
  \centering
  \includegraphics[width=\textwidth]{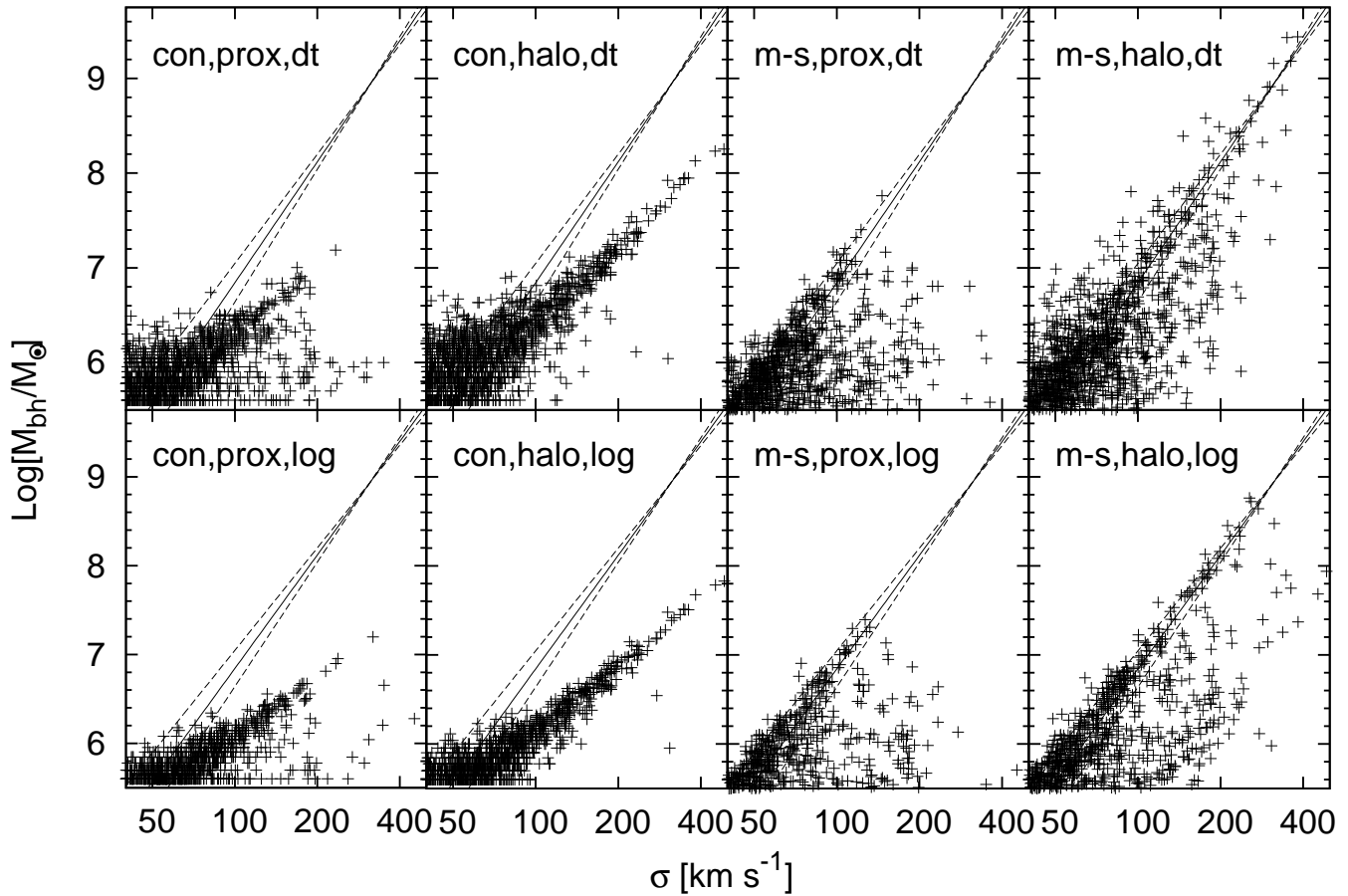}
  \caption{Comparison of models (shown as points) 
           against the observed $\msig$ relation at $z=0$ (solid line). 
           Dashed lines indicate $1\sigma$ uncertainty bounds 
           in the observed relation.}
\label{fig:msigma}
\end{figure*}

\begin{figure*}
  \centering
  \includegraphics[width=\textwidth]{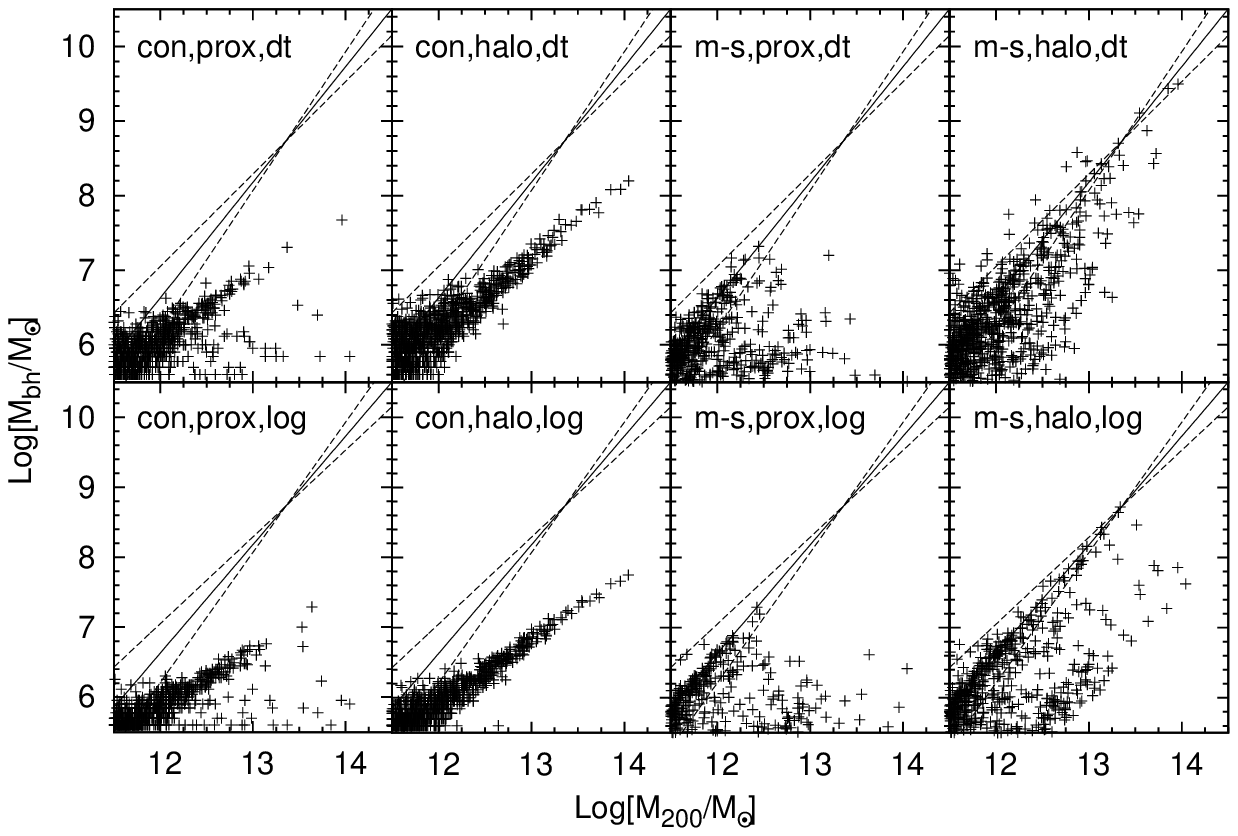}
  \caption{Comparison of models (shown as points) against the observed $\mhalo$ 
           relation at $z=0.2$ (solid line). Uncertainties are not shown 
           but are included in later analysis.
           Dashed lines indicate $1\sigma$ uncertainty bounds in 
           the observed relation.}
\label{fig:mhalo}
\end{figure*}

We compare the results of our simulations to the known $\msig$ relation 
at $z=0$ (Eq.~\ref{eq:msigma}) in Figure~\ref{fig:msigma} 
and to the observed $\mhalo$ relation 
at $z=0.2$ (Eq.~\ref{eq:mhalo}) in Figure~\ref{fig:mhalo}. 
We have arranged these plots such that the most ``pessimistic'' 
combinations of models --- constant initial mass, distance and 
velocity tests for mergers, and new halo and merger checks 
evenly spaced in $\log a$ --- are located in the lower-left portions of the 
plots, while the most ``optimistic'' scenarios are in the upper right.
Our choice of model can greatly affect a number of aspects 
of SMBH relations, including the scaling of SMBH mass 
with $\sigma$ and ${M}_{\rm tot}$, the maximum mass 
of an SMBH, and the amount of scatter in the produced relations.

\begin{figure}
  \centering
  \includegraphics[width=\columnwidth]{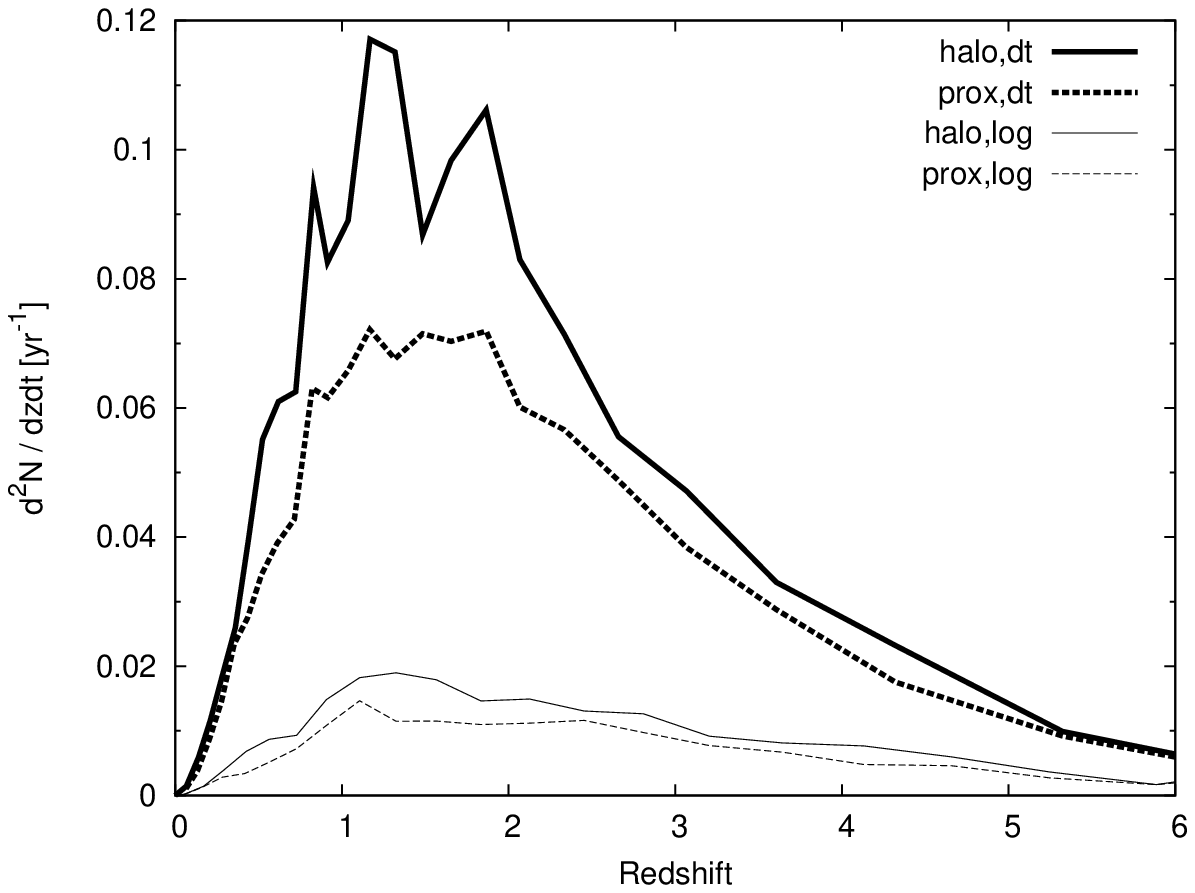}
  \caption{Merger rate of all SMBHs in the simulation volume as a
           function of redshift. We only show models with constant initial 
           mass. Models with instant halo-based merging are shown with 
           solid lines, and models with merging tests are shown with
           dotted lines. Bold lines describe models with merger 
           checks at every time step; thin lines show models with 
           infrequent checks.}
\label{fig:bhmerge}
\end{figure}

Models with instant SMBH mergers produce relations with somewhat lower 
scatter, especially at higher masses, 
and generated higher maximum black hole mass than models 
with distance and velocity checks for mergers. For ``prox'' models, larger 
halos may contain several unmerged SMBHs, and spurious halo mergers 
do not lead to mergers of the black holes. These cause even the most 
massive halos to host relatively small black holes.
Additionally, the difference in the total merger rate 
can be very dramatic, as shown in Figure~\ref{fig:bhmerge}.
The merger rate is defined as
\begin{equation}
  \frac{d^2 N}{dz dt} \approx \frac{\Delta N}{\Delta z \Delta V}
                         4 \pi c (1+z)^2 d_A^2(z),
\label{eq:merge}
\end{equation} 
where $\Delta N$ is the number of mergers in the redshift interval 
$\Delta z$, $\Delta V$ is our simulated volume, $d_A(z)$ 
is the angular diameter distance, and $c$ is the speed of light.
The difference between models is 
especially significant at low redshift, where ``prox'' models can reduce 
the peak merger rate by a factor of two. The difference is 
negligible at high redshift, since
the merger rate here is driven mostly by collisions of smaller halos, and 
the differences between ``prox''- and ``halo''-based merging are 
the smallest.
For all cases, our measured merger rate is less than rates 
found by merger trees 
(e.g.,~\citet{micic_black_2008};~\citet{menou_merger_2001}) 
since those works can include SMBH masses 
below those which we can resolve.

The frequency of merger checks can have a moderate impact on the 
slope of the final relations and a significant impact on 
the maximum black hole mass. 
However, changing this portion of the model does not significantly 
alter the scatter for models with constant initial mass. Checking 
for mergers every time step raises the maximum SMBH mass by roughly 
half an order of magnitude and allows smaller halos ($< 10^{12} \msol$) 
to host black holes fitting the observed relations. We can explain 
these results by studying the merger rate in Figure~\ref{fig:bhmerge}.
Here, we see that merger checks at every time step can increase the merger 
rate by an order of magnitude relative to merger checks evenly 
spaced in $\log a$. This is largely due to an increased number of 
seed black holes: by checking at every time step, we may capture halos 
as soon as they become resolvable, and before they encounter their 
first merger event. Indeed, we seed roughly twice as many SMBHs in the 
``dt'' models relative to the ``log'' models, even though they both 
end up with roughly the same number of black holes at $z=0$. 
Also, we capture more mergers at late times: in ``log'' models, 
we may skip the formation of small halos and their merging 
onto an already-formed larger neighbor, missing the SMBH mass 
associated with the smaller halo. This occurs regardless of 
seeding method.
Spurious re-seeding also contributes somewhat, but
this does not dominate because we see the same relative numbers 
of seeded and final black holes in both the ``prox'' and ``halo'' 
models. 

\begin{figure}
  \centering
  \includegraphics[width=\columnwidth]{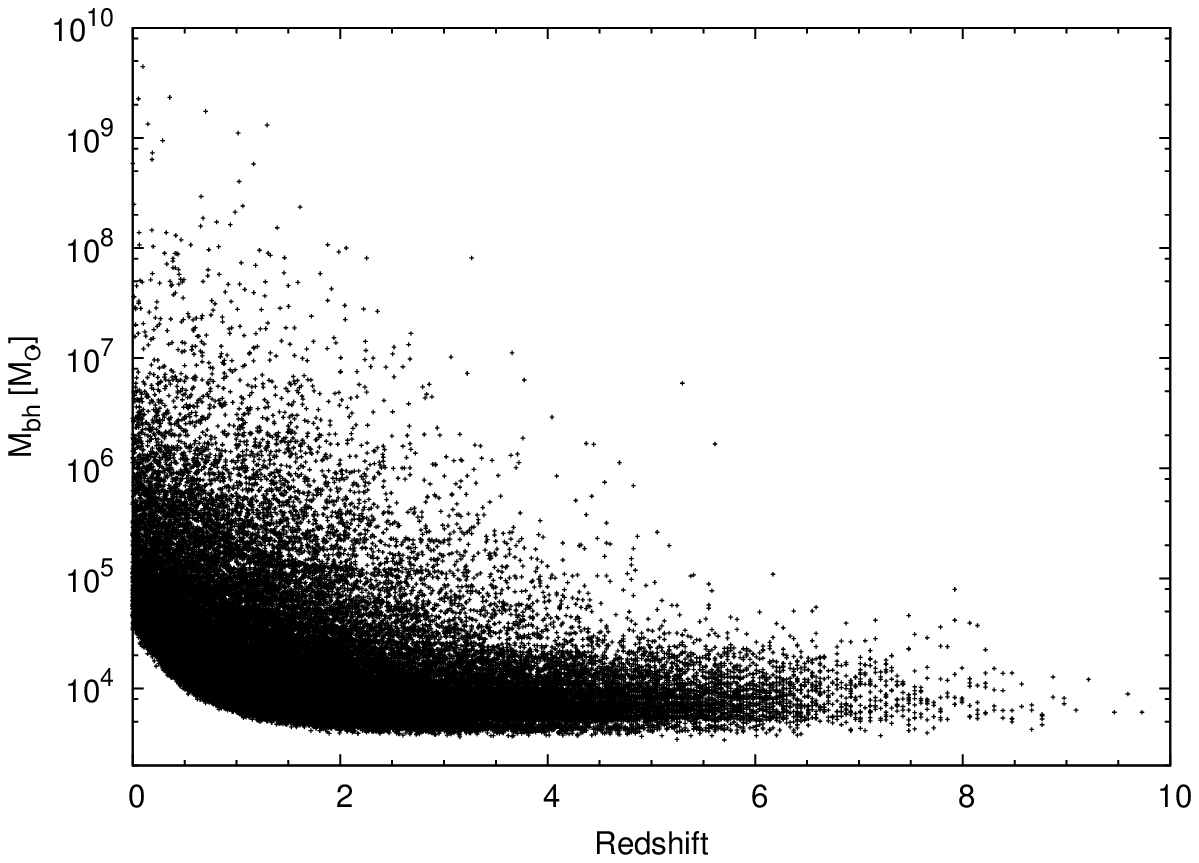}
  \caption{Initial SMBH mass for seed black holes as a function of redshift 
           for the ``m-s,halo,dt'' run.}
\label{fig:bhinit}
\end{figure}

Using an initial seed based on early $\msig$ relations produces 
broad scatter in the final relations. This is especially evident 
in the ``m-s,halo,log'' combination, in which there appear to be 
two distinct populations of SMBHs: one population along the observed 
relation and another at lower mass. This behavior is due to the 
fact that halos are seeded in two scenarios: when the halo first 
becomes resolvable in the simulation, and if the halo merges 
with another halo, loses its black hole, and later separates. 
Figure~\ref{fig:bhinit} illustrates this by showing the 
initial seed mass as a function of redshift. Note that seed black 
holes are never larger than $\sim 10^{-5}$ of the host halo mass.
Small halos continue to appear throughout the evolution of the 
simulation and are seeded with $\sim 10^4 \msol$ black holes. However, 
occasionally a larger halo loses its black hole and must be 
re-seeded with a correspondingly larger SMBH. Thus we may be left 
with two populations: halos with their original SMBHs that evolve 
to relations similar to halos with constant initial mass, and halos 
that are re-seeded at lower redshifts with black hole masses 
closer to the observed relation. This distinction is largely eliminated
by applying the merger tests at every time step and merging 
black holes instantly 
on halo mergers. This results both in higher mass due to increased mergers 
and in increased low-redshift re-seeding of high-mass halos. This 
implies that re-seeding selects a few high-mass halos and places them 
on the observed relation at late times. Indeed, we find that all
of the SMBHs with $\mbh > 10^9 \msol$ and those black holes that lie at 
or above the observed relations 
(roughly half of the population at intermediate masses) 
are the product of low-redshift 
re-seeding. However, the majority of black holes evolve without 
re-seeding at late times, so that the differences between ``m-s,dt'' 
and ``m-s,log'' models at intermediate masses are largely due to the 
increased merger rate promoting the masses of all black holes. 
Note that the earlier discovery of halos by checking at every time step 
does not play a significant role here, since halos discovered later 
are simply initialized with larger black holes (which was one of the objects 
of this model).

\begin{figure}
  \centering
  \includegraphics[width=\columnwidth]{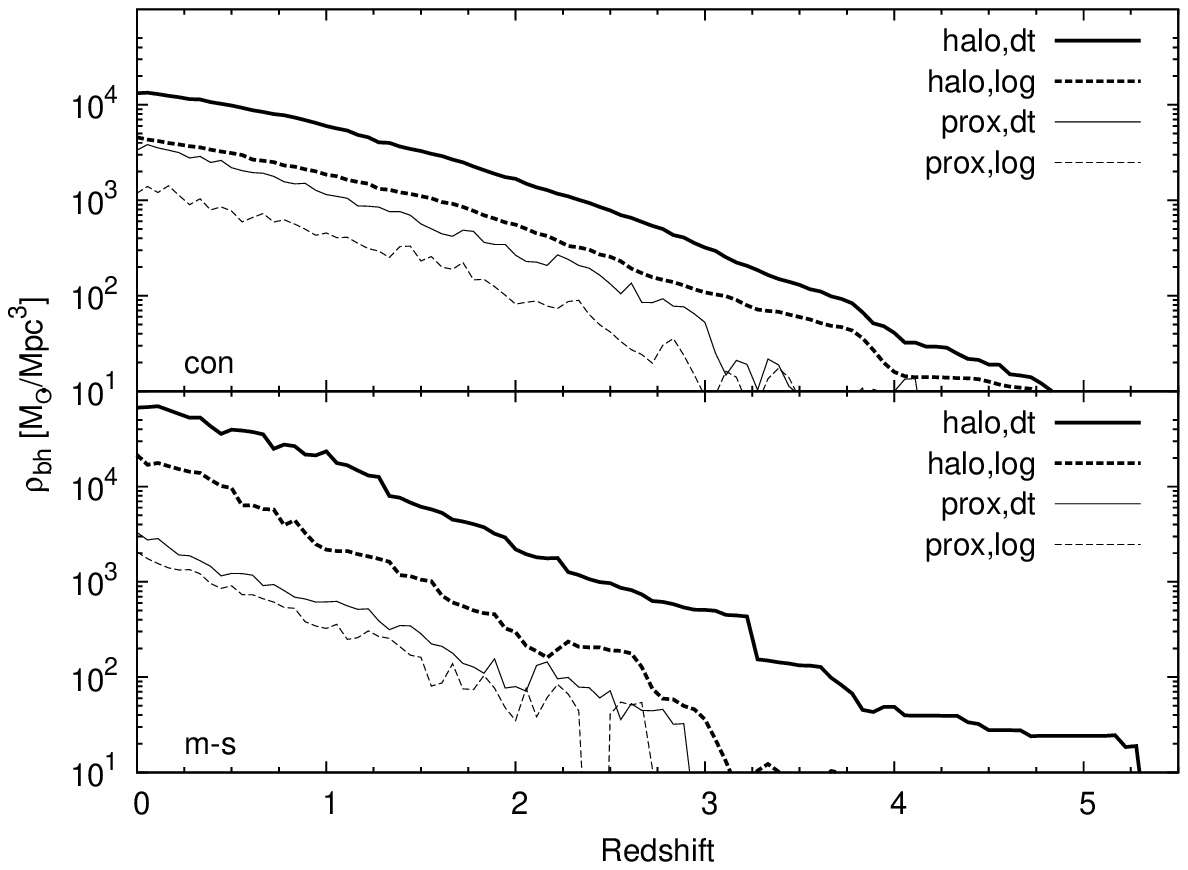}
  \caption{Cosmic density of SMBHs with masses $10^6 \msol < {M}_{\rm bh} < 5
           \times 10^{9} \msol$ as a function of redshift. The top panel shows
           all models with constant initial seed mass, while the bottom panel 
           shows models with seed masses based on the $\msig$ relation.
           Models with instant merging are shown with bold lines; those 
           with distance and velocity tests for merging are 
           shown with thin lines. 
           Finally, solid lines 
           describe models with merger checks at every time step, and 
           dashed lines indicate models with checks evenly spaced in 
           $\log a$.}
\label{fig:bhden}
\end{figure}

We may also understand these results by examining the cosmic
SMBH mass density, as in Figure~\ref{fig:bhden}. Note that
following~\citet{shankar_supermassive_2004} we only count 
black holes that are matched to a halo and have 
masses $10^6 < {M}_{\rm bh} / \msol < 5 \times 10^{9}$ when computing the
mass density. All the SMBH densities produced by our models are well 
below the observed $z=0$ density of 
$4.3 \times 10^5 \msol {\rm Mpc}^{-3}$~\citep{shankar_supermassive_2004}. 
Despite the fact that the smallest halos in the ``m-s'' runs started with 
 $10^4 \msol$ seed black holes, re-seeding and merging at 
low redshift causes the largest black holes to reach 
$\sim 10^9 \msol$ and the cosmic mass density to reach nearly $1/10$ 
of the observed value.

Even though the re-seeding artificially creates all the SMBHs 
larger than $10^9 \msol$, and half the black holes above $10^7 \msol$, 
the majority of black holes are not re-seeded and have masses lower 
than that of the relation. Thus, even though we can reproduce the 
m-sigma relation at high masses (albeit with high scatter) there are 
not enough black holes with sufficient mass to reach the observed 
cosmic mass density.
Similarly, the higher merger rates produced by 
instant merging models promote more black holes above the minimum 
mass threshold for inclusion in the density calculation. 
The frequency of merger 
checks has a dramatic impact on the $z=0$ density: here there is 
up to a factor of 
five difference between models. This is largely due to the increased 
rate of discovering early halos in models with constant 
initial mass, thereby adding more SMBH mass 
to the simulation at early times. However, the extra mass added 
when checking ``m-s'' models at every time step  
is largely due to increased re-seeding.

We quantify our comparisons by fitting our resulting $\msig$ and $\mhalo$ 
relations to a straight line in log space. Since the ``prox'' models 
are obviously poor fits to power laws, we will not 
include them. To mimic observations, 
we only include black holes 
with masses greater than $10^6 \msol$. Tables~\ref{tab:msigma} 
and~\ref{tab:mhalo} describe the differences between fits 
to our models and the observed relations. 
We assume the halo mass uncertainty of Eq.~(\ref{eq:deltaM}); however,
since it is 
difficult to quantify all the uncertainties for $\sigma$ 
generated in cosmological simulations, 
we will ignore them in computing fits. While this is an admittedly 
crude procedure, it does give us some estimate of the 
ability to distinguish these models from observations and 
from each other.  
We define the error in the tables to be the difference between 
the result of the model in the simulation and the observed quantity 
divided by the sum of their respective $1\sigma$ uncertainties.
\begin{table}
  \centering
  \caption{Best-fit Slope and Normalization with $1\sigma$ Uncertainty, 
           Compared to Observed $\msig$ Relation at $z=0$ 
           (Eq.~\ref{eq:msigma}).}
  \begin{tabular}{ccccc}
\hline
\hline
 Model & Slope & Error & Normalization & Error \\
\hline
con,halo,log &  2.2 $\pm$  0.1 & -4.4 &  6.9 $\pm$  0.3 & -3.2 \\
con,halo,dt &  1.9 $\pm$  0.1 & -5.1 &  7.3 $\pm$  0.4 & -2.0 \\
m-s,halo,log &  4.6 $\pm$  0.3 &  0.6 &  8.0 $\pm$  1.0 & -0.2 \\
m-s,halo,dt &  4.1 $\pm$  0.2 & -0.3 &  8.1 $\pm$  0.5 & -0.1 \\
\hline
\end{tabular}

  \tablecomments{Error is defined in the text.}
  \label{tab:msigma}
\end{table}
\begin{table}
  \centering
  \caption{Best-fit Slope and Normalization with $1\sigma$ Uncertainty, 
           Compared to Observed $\mhalo$ Relation at $z=0.2$ 
           (Eq.~\ref{eq:mhalo}).}
  \begin{tabular}{ccccc}
\hline
\hline
 Model & Slope & Error & Normalization & Error \\
\hline
con,halo,log &  0.89 $\pm$  0.01 &  -2.0 &  6.8 $\pm$  0.2 & -4.9 \\
con,halo,dt &  0.87 $\pm$  0.01 &  -2.1 &  7.2 $\pm$  0.2 & -3.4 \\
m-s,halo,log &  2.1 $\pm$  0.2 &   1.0 &  7.0 $\pm$  1.0 & -1.0 \\
m-s,halo,dt &  1.70 $\pm$  0.05 &   0.4 &  7.9 $\pm$  0.4 & -0.6 \\
\hline
\end{tabular}

  \tablecomments{Error is defined in the text.}
  \label{tab:mhalo}
\end{table}

We see that, in general, models with constant initial mass are 
indistinguishable from each other, while these models are, as a 
group, significantly different from models with varying initial mass. 
None of the constant-mass models produce enough black hole mass to 
match the observed relations, as 
seen in the low normalization values, and they do not produce enough high-mass 
SMBHs in the largest halos, as seen in the low slope. 

In contrast, seeding halos with black holes with masses from an early 
$\msig$ relation matches observations well at low redshift. 
While these models fit the observed relations well, the total 
mass density of SMBHs is still well below the observed value. 

Surprisingly, we see marginally steeper slopes 
in the ``log'' models than in the ``dt'' 
models. Even though ``dt'' models have higher merger rates 
and produce more massive black holes, they also generate a large 
population of moderate-mass black holes hosted in small halos.
This skews the fits to flatter slopes despite the higher maximum 
black hole mass. However, these differences are not statistically 
significant, except for the constant initial mass combined with 
instant merging scenarios, where the low scatter produces small 
uncertainties.

Generally, the difference between a particular model and the $\msig$ 
relation is matched by the difference between that same model and 
the $\mhalo$ relation. However, the models are more indistinguishable 
from the 
$\mhalo$ relation, mostly due to the larger uncertainties 
in the observed relation. 


\section{Conclusion}
We have developed a new fast, parallel halo finder for inclusion
in cosmological simulations with the simulation code FLASH.
Using SO halo finding techniques, we are 
able to produce halo catalogs in good agreement with 
traditional post-processing halo finders. Since our halo finder 
is designed to be fast, we are able to perform halo finding 
operations at every time step in the simulation, allowing 
us to perform a detailed analysis of SMBH subgrid models.

While merging alone cannot generate enough total mass in SMBHs 
to match the observed cosmic mass density or generate high enough 
maximum black hole mass in the largest halos to match the 
observed $\msig$ and $\mhalo$ relations, it 
can play a large role in developing the slope of the relations, 
especially at intermediate mass ranges. Thus, merging should 
not be totally discounted in considering the processes that 
provide the correlations between black hole mass and bulge, galaxy, 
and halo properties.
However, since none of our considered models can account for the observed 
cosmic mass density, there is still a significant role for 
accretion and feedback processes in the evolution of SMBHs.
Also, since the choice of models can greatly affect the 
cosmic SMBH mass density, accretion and feedback models must be 
chosen carefully to match observations.

The choice of subgrid models can dramatically impact the 
merging rate of black holes. Since the merger rate has a large 
influence on the performance of upcoming gravitational 
wave detectors, halo finding operations in simulations 
should be done as frequently as possible in order to accurately 
capture this rate. 
We do not believe that the inclusion of gas will significantly 
affect these relative differences, since they are largely driven 
by the ability to find more black holes at early times.
However, the choice of merging test does 
not greatly affect the predicted rate, except at low redshift.

While we have bracketed the possible subgrid models with 
``optimistic'' and ``pessimistic'' scenarios, models 
best matching insights from theory and observations are usually 
in between those extremes. While seeding black holes with 
a uniform initial mass for black holes may well model 
high-redshift behavior, it is not clear that this is a 
useful strategy for re-seeding low-redshift halos. 
Most re-seeding is certainly an artifact of the halo finder 
and the lack of an in-code merger tree to determine when halos 
have truly merged. However, there are some plausible scenarios 
where re-seeding may be needed: for example, when three-body or 
gas interactions strip an SMBH from a central galaxy.
Also, when this
approach is coupled with infrequent halo finding operations, it 
may deposit too little mass in the seed SMBHs. While instant 
merging is too optimistic, our current lack of understanding of 
SMBH mergers implies that we cannot entirely specify this portion 
of the subgrid model, and we must rely on a bracketing procedure. 

Future examination of these subgrid models must be done in a cosmological 
simulation involving gas evolution, accretion onto the black holes, 
and feedback from active galactic nuclei. 
However, since models of these processes 
carry with them their 
own assumptions and adjustable parameters, care must be taken to 
fully separate the effects of black hole seeding and merging. 
It may be possible that due to the self-regulating nature of feedback 
that the differences among these models may disappear; however, 
the differences in merger rate and peak black hole mass suggest 
significant variances may remain.
Only 
once all aspects of these subgrid models are analyzed, understood, and 
compared to our observational understanding can we confidently combine them 
into an integrated model.
 
\section*{Acknowledgments}
The authors acknowledge support under a Presidential Early 
Career Award from the U.S. Department of Energy, 
Lawrence Livermore National Laboratory (contract B532720).
Additional support was provided by a DOE 
Computational Science Graduate Fellowship 
(DE-FG02-97ER25308) and the National Center for 
Supercomputing Applications.
The software used in this work was in part developed by the DOE-supported 
ASC/Alliance Center for Astrophysical Thermonuclear Flashes 
at the University of Chicago.
This research used resources of the National Center for Computational 
Sciences at Oak Ridge National Laboratory, which is supported by the 
Office of Science of the US Department of Energy under 
contract no. DE-AC05-00OR22725. The authors thank 
Brian O'Shea, Tiziana Di Matteo, and many others for useful 
conversations at the tenth Great Lakes Cosmology Workshop.

\appendix
\label{sec:scaling}
\section{Performance and Parallel Scalability}
In order to minimize communication among processors, 
we divide potential halos into 
two lists. For every potential halo, 
we compute the overdensity at the largest possible 
on-processor radius. If the next search radius is larger than this, the halo is 
added to a list that must be communicated. Thus, all on-processor 
halos are processed concurrently with no communication. Since the 
volume of communication is very small, we send all other halos 
to every processor. However, since this approach is not scalable to very 
large halo catalogs or numbers of processors, we switch 
to a buffered communication pattern if we cannot allocate 
a global halo catalog. In the buffered approach, halo lists 
are only communicated to the nearest processors one at a time. 
The list is communicated until all blocks within the volume of the 
processor's halos 
have been searched. 
While slower than the all-to-all approach, the amount of storage 
per processor required stays fixed as the number of processors and 
number of halos grow. In either case, communication continues until 
all halos are fully searched. 
For this study, we will use the all-to-all approach. 

\begin{figure}
  \centering
  \subfigure[Strong scaling of pSO halo finder at various problem sizes as
               a fraction of total wall clock time in a single FLASH time step.]
           { \includegraphics[scale=0.67]{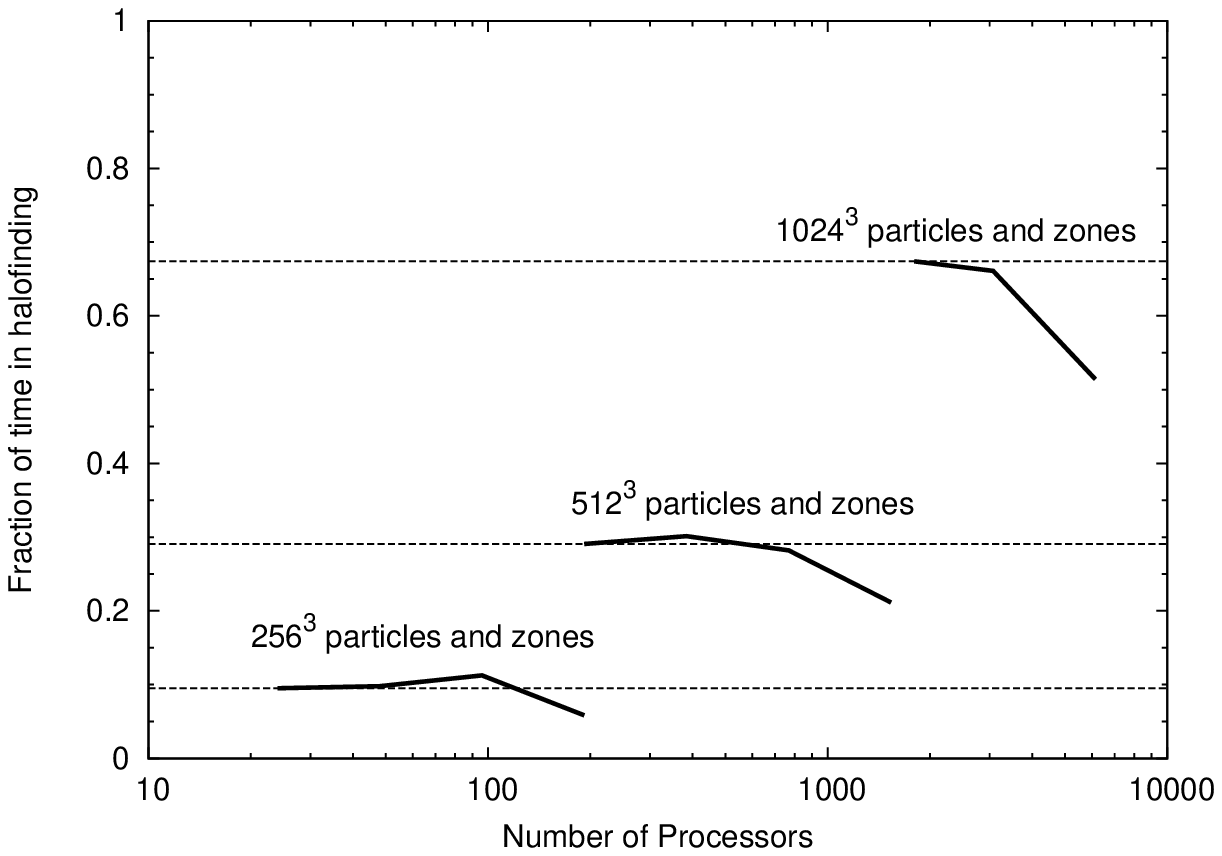}
             \label{fig:scaling} }
  \subfigure[Weak scaling at smallest number of processors for each 
               problem size as a fraction of total wall clock time in a 
               single FLASH time step. The dotted line indicates weak scaling 
               after adjustments have been made to account for 
               non-linear scaling of halo counts with problem size.]
           { \includegraphics[scale=0.67]{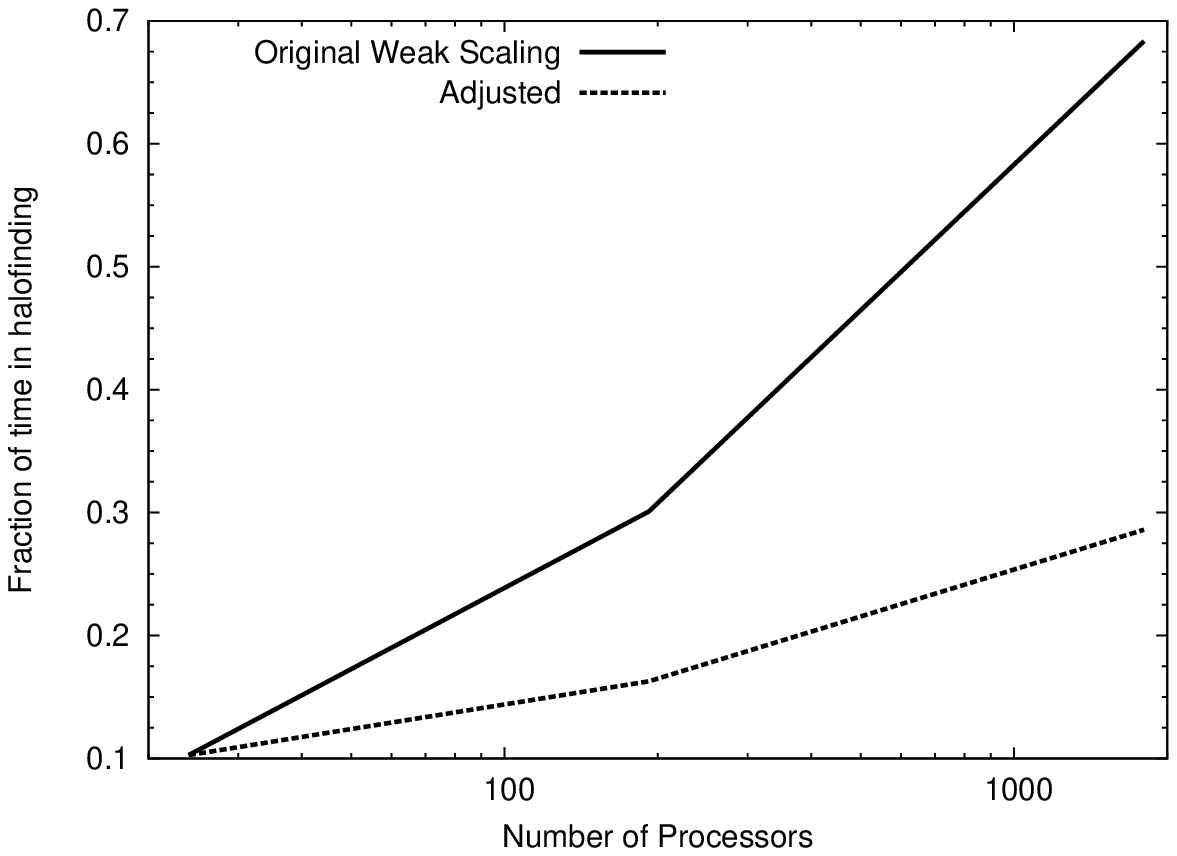}
             \label{fig:adjusted} }
  \caption{Strong and weak scaling of the pSO halo finder.}
\end{figure}

Figure~\ref{fig:scaling} shows the strong scaling of the pSO halo finder 
as a fraction of the total wall clock time in a single FLASH time step. 
We show three different uniform problem sizes: $256^3$, $512^3$, and $1024^3$ 
particles and zones. We restarted each simulation at a representative redshift, 
$z=0.25$, and ran for 
five time steps. These times do not include optional portions of the halo 
finder routine, such as writing the halo catalog to disk or tagging particles 
within halos. We performed these calculations on jaguar, a Cray XT5 
system at Oak Ridge National Laboratory. Jaguar consists 
of 16,688 dual six-core AMD Opteron nodes with 16 GB of memory 
per node and has a peak performance of 2.332 petaflops.
Our approach offers good strong scaling behavior: 
we are able to beat or match the scaling performance of FLASH at 
all problem sizes. 
At larger core counts, FLASH has difficulty scaling the Poisson 
solver, whereas the halo finder maintains good scalability.
However, 
we can infer that we have poor weak scaling: the halo finding steps 
require ever larger wall clock time as the problem size grows.
 
The poor weak scaling is due to several factors. First, since these are 
uniform grid calculations, at low redshift the particle distribution among 
processors becomes highly unbalanced. FLASH is block based and uses 
a Morton curve the distribute these blocks among the processors. Thus, 
while each processor has roughly the same number of blocks, those 
blocks in highly dense regions will contain many more particles 
than those in voids. Since our halo finder scans through particles, 
there is a lack of concurrency due to this imbalance. At small 
problem sizes, this is not an issue, but with $1024^3$ particles all 
processors must wait for the few heavily loaded processors to complete 
searching. There is an imbalance in halos as well, 
since highly overdense regions 
contain more halos than underdense regions.
Note that this is a bigger problem for box sizes smaller than 
$200$~${\rm Mpc}$.
We found that the $256^3$ run placed a maximum of $15$ halos 
on a single processor, the $512^3$ run had up to $33$ halos on a single 
processor, and the $1024^3$ run mapped up to $80$ halos on a single 
processor. Figure~\ref{fig:adjusted} shows the effects of the 
imbalanced load due to an increased maximum number of halos on the weak 
scaling behavior. We compute 
the effects by maintaining no more than $15$ halos per processor no matter 
the problem size. We performed this test by ignoring halos past this limit. 
We see that halo imbalance explains a significant 
part of the weak scaling results.

Second, as we increase the resolution, we greatly increase the number of 
resolvable halos. 
Since smaller halos are much more numerous than larger ones (see 
Figure~\ref{fig:massFunc}), for every doubling of the resolution 
we get roughly ten times the number of halos. However, the number of 
particles and 
zones only increases by a factor of eight. This means that even if 
we devote eight times the number of processors to the problem, we will spend a
 larger fraction of a time step finding all the halos. 

Finally, as we increase the resolution, we require more steps in the 
binary search procedure to meet the stopping criterion. At $256^3$ zones, 
we require roughly $35$ steps to identify the halos. At $512^3$ zones, 
we require $45$ steps on average. Finally, for $1024^3$ zones, we 
average $55$ steps. This increases the serial runtime of the halo finder, 
regardless of the degree of concurrency.

These calculations were performed with only dark matter. As we perform 
runs with more included physics, such as hydrodynamics and radiative 
cooling, the fraction of time spent in halo finding will decrease, 
even for high-resolution runs. Also, we expect AMR runs to alleviate 
the problems of poor particle load balancing, since in cosmological 
simulations we typically refine on overdense regions. We are 
currently preparing a manuscript detailing the performance of the halo finder 
with different box sizes, combinations of resolutions of particles 
and zones, additional physics, and different AMR refinement schemes.
\bibliography{ms}		

\begin{thebibliography}{54}
\expandafter\ifx\csname natexlab\endcsname\relax\def\natexlab#1{#1}\fi

\bibitem[{{Amaro-Seoane} {et~al.}(2007){Amaro-Seoane}, Gair, Freitag, Miller,
  Mandel, Cutler, \& Babak}]{amaro-seoane_topical_2007}
{Amaro-Seoane}, P., Gair, J.~R., Freitag, M., Miller, M.~C., Mandel, I.,
  Cutler, C.~J., \& Babak, S. 2007, Class. Quantum Grav., 24, 113

\bibitem[{Baes {et~al.}(2003)Baes, Buyle, Hau, \&
  Dejonghe}]{baes_observational_2003}
Baes, M., Buyle, P., Hau, G. K.~T., \& Dejonghe, H. 2003, \mnras, 341, L44

\bibitem[{Baker {et~al.}(2008)Baker, Boggs, Centrella, Kelly, {McWilliams},
  Miller, \& van Meter}]{baker_modeling_2008}
Baker, J.~G., Boggs, W.~D., Centrella, J., Kelly, B.~J., {McWilliams}, S.~T.,
  Miller, M.~C., \& van Meter, J.~R. 2008, \apjl, 682, L29

\bibitem[{Bandara {et~al.}(2009)Bandara, Crampton, \&
  Simard}]{bandara_relationship_2009}
Bandara, K., Crampton, D., \& Simard, L. 2009, \apj, 704, 1135, in press
  (arXiv:0909.0269)

\bibitem[{Baugh(2006)}]{baugh_primerhierarchical_2006}
Baugh, C.~M. 2006, Rep. Prog. in Phys., 69, 3101

\bibitem[{Begelman {et~al.}(2006)Begelman, Volonteri, \&
  Rees}]{begelman_formation_2006}
Begelman, M.~C., Volonteri, M., \& Rees, M.~J. 2006, \mnras, 370, 289

\bibitem[{Berczik {et~al.}(2006)Berczik, Merritt, Spurzem, \&
  Bischof}]{berczik_efficient_2006}
Berczik, P., Merritt, D., Spurzem, R., \& Bischof, H.~P. 2006, \apjl, 642, L21

\bibitem[{Bertschinger(2001)}]{bertschinger_multiscale_2001}
Bertschinger, E. 2001, \apjs, 137, 1

\bibitem[{{Bhattacharya} {et~al.}(2010){Bhattacharya}, {Heitmann}, {White},
  {Luki{\'c}}, {Wagner}, \& {Habib}}]{bhattacharya_mass_2010}
{Bhattacharya}, S., {Heitmann}, K., {White}, M., {Luki{\'c}}, Z., {Wagner}, C.,
  \& {Habib}, S. 2010, arXiv e-prints, arXiv:1005.2239

\bibitem[{Bogdanovic {et~al.}(2007)Bogdanovic, Reynolds, \&
  Miller}]{bogdanovic_alignment_2007}
Bogdanovic, T., Reynolds, C.~S., \& Miller, M.~C. 2007, \apjl, 661, L147

\bibitem[{Booth \& Schaye(2009)}]{booth_cosmological_2009}
Booth, C.~M. \& Schaye, J. 2009, \mnras, 398, 53–74

\bibitem[{Br\"{u}ggen \& Scannapieco(2009)}]{brggen_self-regulation_2009}
Br\"{u}ggen, M. \& Scannapieco, E. 2009, \mnras, 398, 548

\bibitem[{Carilli \& Taylor(2002)}]{carilli_c_2002}
Carilli, C.~L. \& Taylor, G.~B. 2002, ARA\&A, 40, 319–348

\bibitem[{Cattaneo \& Teyssier(2007)}]{cattaneo_agn_2007}
Cattaneo, A. \& Teyssier, R. 2007, \mnras, 376, 1547

\bibitem[{Croton(2009)}]{croton_simple_2009}
Croton, D.~J. 2009, \mnras, 394, 1109

\bibitem[{{Di Matteo} {et~al.}(2008){Di Matteo}, Colberg, Springel, Hernquist,
  \& Sijacki}]{di_matteo_direct_2008}
{Di Matteo}, T., Colberg, J., Springel, V., Hernquist, L., \& Sijacki, D. 2008,
  \apj, 676, 33

\bibitem[{Dotti {et~al.}(2007)Dotti, Colpi, Haardt, \&
  Mayer}]{dotti_supermassive_2007}
Dotti, M., Colpi, M., Haardt, F., \& Mayer, L. 2007, \mnras, 379, 956

\bibitem[{Evrard {et~al.}(1996)Evrard, Metzler, \& Navarro}]{evrard_mass_1996}
Evrard, A.~E., Metzler, C.~A., \& Navarro, J.~F. 1996, \apj, 469, 494

\bibitem[{Evrard {et~al.}(2008)}]{evrard_virial_2008}
Evrard, A.~E. {et~al.} 2008, \apj, 672, 122

\bibitem[{Fan(2006)}]{fan_evolution_2006}
Fan, X. 2006, New Astron. Rev., 50, 665

\bibitem[{Ferrarese(2002)}]{ferrarese_beyondbulge:fundamental_2002}
Ferrarese, L. 2002, \apj, 578, 90

\bibitem[{Fryxell {et~al.}(2000)}]{fryxell_flash:adaptive_2000}
Fryxell, B. {et~al.} 2000, ApJS, 131, 273

\bibitem[{{Graham}(2007)}]{graham_black_2007}
{Graham}, A.~W. 2007, \mnras, 379, 711

\bibitem[{Gu {et~al.}(2009)Gu, Cao, \& Jiang}]{gu_bulk_2009}
Gu, M., Cao, X., \& Jiang, D.~R. 2009, \mnras, 396, 984–996

\bibitem[{Gultekin {et~al.}(2009)}]{gultekin_m-sigma_2009}
Gultekin, K. {et~al.} 2009, \apj, 698, 198

\bibitem[{Hockney \& Eastwood(1988)}]{hockney_computer_1988}
Hockney, R.~W. \& Eastwood, J.~W. 1988, Computer Simulation using Particles
  (Bristol: Hilger)

\bibitem[{Hopkins {et~al.}(2005)Hopkins, Hernquist, Cox, Matteo, Martini,
  Robertson, \& Springel}]{hopkins_black_2005}
Hopkins, P.~F., Hernquist, L., Cox, T.~J., Matteo, T.~D., Martini, P.,
  Robertson, B., \& Springel, V. 2005, \apj, 630, 705

\bibitem[{Hopkins {et~al.}(2010)Hopkins, Younger, Hayward, Narayanan, \&
  Hernquist}]{hopkins_mergers_2010}
Hopkins, P.~F., Younger, J.~D., Hayward, C.~C., Narayanan, D., \& Hernquist, L.
  2010, \mnras, 402, 1693

\bibitem[{Islam {et~al.}(2004)Islam, Taylor, \& Silk}]{islam_massive_2004}
Islam, R.~R., Taylor, J.~E., \& Silk, J. 2004, \mnras, 354, 427

\bibitem[{Koushiappas {et~al.}(2004)Koushiappas, Bullock, \&
  Dekel}]{koushiappas_massive_2004}
Koushiappas, S.~M., Bullock, J.~S., \& Dekel, A. 2004, \mnras, 354, 292

\bibitem[{Luki\'{c} {et~al.}(2007)Luki\'{c}, Heitmann, Habib, Bashinsky, \&
  Ricker}]{lukic_halo_2007}
Luki\'{c}, Z., Heitmann, K., Habib, S., Bashinsky, S., \& Ricker, P.~M. 2007,
  \apj, 671, 1160

\bibitem[{Luki\'{c} {et~al.}(2009)Luki\'{c}, Reed, Habib, \&
  Heitmann}]{lukic_structure_2009}
Luki\'{c}, Z., Reed, D., Habib, S., \& Heitmann, K. 2009, \apj, 692, 217

\bibitem[{Madau \& Rees(2001)}]{madau_massive_2001}
Madau, P. \& Rees, M.~J. 2001, \apjl, 551, L27

\bibitem[{Magorrian {et~al.}(1998)}]{magorrian_demography_1998}
Magorrian, J. {et~al.} 1998, \aj, 115, 2285

\bibitem[{Mayer {et~al.}(2007)Mayer, Kazantzidis, Madau, Colpi, Quinn, \&
  Wadsley}]{mayer_rapid_2007}
Mayer, L., Kazantzidis, S., Madau, P., Colpi, M., Quinn, T., \& Wadsley, J.
  2007, Science, 316, 1874

\bibitem[{Menou {et~al.}(2001)Menou, Haiman, \& Narayanan}]{menou_merger_2001}
Menou, K., Haiman, Z., \& Narayanan, V.~K. 2001, \apj, 558, 535

\bibitem[{Merloni {et~al.}(2010)}]{merloni_cosmic_2010}
Merloni, A. {et~al.} 2010, \apj, 708, 137

\bibitem[{Merritt \& Ferrarese(2001)}]{merritt_relationship_2001}
Merritt, D. \& Ferrarese, L. 2001, in ASP Conf. Ser. 249, Vol. 249, The Central
  Kiloparsec of Starbursts and {AGN:} The La Palma Connection, ed. J.~E. B.
  J.~H. Knapen, 335

\bibitem[{Merritt \& Milosavljevic(2005)}]{merritt_massive_2005}
Merritt, D. \& Milosavljevic, M. 2005, Liv. Rev. Rel., 8, 8

\bibitem[{Micic {et~al.}(2008)Micic, {Holley-Bockelmann}, \&
  Sigurdsson}]{micic_black_2008}
Micic, M., {Holley-Bockelmann}, K., \& Sigurdsson, S. 2008, \mnras (submitted),
  arXiv:0805.3154

\bibitem[{Micic {et~al.}(2007)Micic, {Holley-Bockelmann}, Sigurdsson, \&
  Abel}]{micic_supermassive_2007}
Micic, M., {Holley-Bockelmann}, K., Sigurdsson, S., \& Abel, T. 2007, \mnras,
  380, 1533

\bibitem[{Ricker(2008)}]{ricker_direct_2008}
Ricker, P.~M. 2008, \apjs, 176, 293

\bibitem[{Ruszkowski {et~al.}(2007)Ruszkowski, Ensslin, Br\"{u}ggen, Heinz, \&
  Pfrommer}]{ruszkowski_impact_2007}
Ruszkowski, M., Ensslin, T.~A., Br\"{u}ggen, M., Heinz, S., \& Pfrommer, C.
  2007, \mnras, 378, 662

\bibitem[{Sesana {et~al.}(2004)Sesana, Haardt, Madau, \&
  Volonteri}]{sesana_lowfrequency_2004}
Sesana, A., Haardt, F., Madau, P., \& Volonteri, M. 2004, \apj, 611, 623

\bibitem[{{Shankar}(2009)}]{shankar_demography_2009}
{Shankar}, F. 2009, New Astron. Rev., 53, 57

\bibitem[{{Shankar} {et~al.}(2004){Shankar}, {Salucci}, {Granato}, {De Zotti},
  \& {Danese}}]{shankar_supermassive_2004}
{Shankar}, F., {Salucci}, P., {Granato}, G.~L., {De Zotti}, G., \& {Danese}, L.
  2004, \mnras, 354, 1020

\bibitem[{Sijacki {et~al.}(2007)Sijacki, Springel, Matteo, \&
  Hernquist}]{sijacki_unified_2007}
Sijacki, D., Springel, V., Matteo, T.~D., \& Hernquist, L. 2007, \mnras, 380,
  877–900

\bibitem[{Tremaine {et~al.}(2002)}]{tremaine_slope_2002}
Tremaine, S. {et~al.} 2002, \apj, 574, 740

\bibitem[{Vernaleo \& Reynolds(2006)}]{vernaleo_agn_2006}
Vernaleo, J.~C. \& Reynolds, C.~S. 2006, \apj, 645, 83

\bibitem[{Volonteri {et~al.}(2003)Volonteri, Haardt, \&
  Madau}]{volonteri_assembly_2003}
Volonteri, M., Haardt, F., \& Madau, P. 2003, \apj, 582, 559

\bibitem[{Volonteri {et~al.}(2008)Volonteri, Lodato, \&
  Natarajan}]{volonteri_evolution_2008}
Volonteri, M., Lodato, G., \& Natarajan, P. 2008, \mnras, 383, 1079

\bibitem[{Warren {et~al.}(2006)Warren, Abazajian, Holz, \&
  Teodoro}]{warren_precision_2006}
Warren, M.~S., Abazajian, K., Holz, D.~E., \& Teodoro, L. 2006, \apj, 646, 881

\bibitem[{White(2002)}]{white_mass_2002}
White, M. 2002, \apjs, 143, 241

\bibitem[{Wise \& Abel(2005)}]{wise_number_2005}
Wise, J.~H. \& Abel, T. 2005, \apj, 629, 615

\end{thebibliography}
\bibliographystyle{apj}	
\nocite{*}

\end{document}